\documentclass[RNAAS]{aastex63}

\usepackage{rotating}
\received{May 15, 2021}
\accepted{October 19, 2021}
\submitjournal{AJ}

\shorttitle{Rotation signature of \textit{TESS}  B-type stars revealed by wavelets}
\shortauthors{Barraza et al.}

\begin{document}

\title{Rotation signature of \textit{TESS}  B-type stars. A comprehensive analysis}

\correspondingauthor{Bruno Leonardo Canto Martins}
\email{brunocanto@fisica.ufrn.br}

\author[0000-0001-5576-8365]{L. F. Barraza}
\affiliation{Departamento de F\'isica Te\'orica e Experimental, Universidade Federal do Rio Grande do Norte, Campus Universit\'ario, Natal, RN, 59072-970, Brazil}
\author[0000-0002-2023-7641]{R. L. Gomes}
\affiliation{Departamento de F\'isica Te\'orica e Experimental, Universidade Federal do Rio Grande do Norte, Campus Universit\'ario, Natal, RN, 59072-970, Brazil}
\author[0000-0002-2425-801X]{Y. S. Messias}
\affiliation{Departamento de F\'isica Te\'orica e Experimental, Universidade Federal do Rio Grande do Norte, Campus Universit\'ario, Natal, RN, 59072-970, Brazil}
\author[0000-0001-5845-947X]{I. C. Le\~ao}
\affiliation{Departamento de F\'isica Te\'orica e Experimental, Universidade Federal do Rio Grande do Norte, Campus Universit\'ario, Natal, RN, 59072-970, Brazil}
\author[0000-0001-5845-947X]{L. A. Almeida}
\affiliation{Escola de Ci\^encia e Tecnologia, Universidade Federal do Rio Grande do Norte, Campus Universit\'ario, Natal, RN, 59072-970, Brazil}
\affiliation{Programa de P\'os-Gradua\c{c}\~ao em F\'isica, Universidade do Estado do Rio Grande do Norte, Mossor\'o, RN, 59610-210, Brazil}
\author[0000-0001-8218-1586]{E. Janot-Pacheco}
\affiliation{Instituto de Astronomia, Geof\'isica e Ci\^encias Atmosf\'ericas, Universidade de S\~ao Paulo, 05508-090, São Paulo, SP, Brazil}
\author[0000-0003-2719-8056]{A. C. Brito}
\affiliation{Instituto Federal do Cear\'a (IFCE), Campus Sobral, Sobral, CE, 62042-030, Brazil}
\author[0000-0002-6642-1057]{F. A. C. Brito}
\affiliation{Instituto Federal de Educa\c{c}\~ao, Ci\^encia e Tecnologia do Cear\'a, Campus Aracati, Aracati - CE, 62800-000, Brazil}
\author[0000-0002-2880-9077]{Santana, J. V.}
\affiliation{Instituto Federal de Educa\c{c}\~ao, Ci\^encia e Tecnologia do Cear\'a, Campus Avan\c{c}ado do Pec\'em, Caucaia - CE, 61680-000, Brazil}
\author[0000-0002-1571-946X]{N. S. Gonçalves }
\affiliation{Instituto Federal do Cear\'a (IFCE), Campus Fortaleza, Av. Treze de Maio, 2081 - Benfica, Fortaleza, CE, 60040-531, Brazil}
\author[0000-0003-1304-6342]{M. L. das Chagas}
\affiliation{Faculdade de F\'isica, Universidade Federal do Sul e Sudeste do Par\'a, 68505-080, Marab\'a PA, Brazil}
\author[0000-0002-5404-8451]{M. A. Teixeira}
\affiliation{Departamento de F\'isica Te\'orica e Experimental, Universidade Federal do Rio Grande do Norte, Campus Universit\'ario, Natal, RN, 59072-970, Brazil}
\author[0000-0001-8218-1586]{J. R. De Medeiros}
\affiliation{Departamento de F\'isica Te\'orica e Experimental, Universidade Federal do Rio Grande do Norte, Campus Universit\'ario, Natal, RN, 59072-970, Brazil}
\author[0000-0001-5578-7400]{B. L. Canto Martins}
\affiliation{Departamento de F\'isica Te\'orica e Experimental, Universidade Federal do Rio Grande do Norte, Campus Universit\'ario, Natal, RN, 59072-970, Brazil}



\begin{abstract}

Stellar rotation is a fundamental observable that drives different aspects of stellar and planetary evolution. In this work, we present an unprecedented manifold analysis of 160 B-type stars with light curves collected by the TESS space mission using three different procedures (Fast Fourier Transform, Lomb-Scargle, and wavelet techniques), accompanied by rigorous visual inspection in the search for rotation periodicities. This effort provides rotational periodicities for 6 new TESS B-type stars and confirmed periodicities for 22 targets with rotation periods previously listed in the literature. For other 61 stars, already classified as possible rotational variables, we identify noisy, pulsational, binarity, or ambiguous variability behavior rather than rotation signatures. The total sample of 28 potential rotators shows an overlap of different classes of rotational variables, composed of $\alpha^2$ Canum Venaticorum, rotating ellipsoidal and SX Arietis stars. The combination of the three techniques applied in our analysis offers a solid path to overcome the challenges in the discrimination of rotation from other variabilities in stellar light curves, such as pulsation, binarity or other effects that have no physical meaning. Finally, the rotational periodicities reported in the present study may represent important constraints for improving stellar evolution  models with rotation, as well as asteroseismic studies of hot stars.

\end{abstract}



\section{Introduction} \label{sec:intro}

Rotation is a fundamental observable for the understanding of the physical mechanisms controlling star path throughout stellar evolution. However, while the detection of periodicities in stellar light curves (LCs) is straightforward, their interpretation in terms of the root causes is a far more challenging task. Indeed, the detection threshold for periodicity depends on the star brightness, its position in the H-R diagram, the time span of observations, and the final post-treatment of the LCs, thus differing from star to star. For instance, the measurement of rotation periodicities in cool stars may be unambiguous due to the presence of quasiperiodic brightness variations in photometric LCs caused by magnetically active regions repeatedly crossing the visible hemisphere as the star rotates (e.g., \citealp{Irwin2009}), but sometimes it may also depend on the applied analysis procedures. For the hottest stars, this scenario may change dramatically due, in particular, to the presence of additional phenomena such as pulsations. For the latter, rotational modulation very likely comes from superficial chemical inhomogeneities, namely, chemical spots, which are typically associated with strong magnetic fields visible at the stellar surface (e.g., \citealp{Michaud1981,Krticka2009,Krticka2012,Krticka2015}). Faced with these features, whereas the current literature lists rotation periodicities for tens of thousands of main-sequence cool stars (e.g., \citealp{Reinhold2013,McQuillan2014,Canto2020}), typically from M- to F-types, only a relatively small number of main-sequence A- and B-type stars are known to exhibit photometric variability that can convincingly be attributed to rotation imprints (e.g., \citealp{Sikora2020}).

The advent of space missions such as CoRoT \citep{Baglin2009}, {\em Kepler} \citep{Borucki2010}, and TESS (\citealp{Ricker2015}) radically changes our perception of stellar variability in different regions of the H-R diagram, where stellar rotation is now being widely revealed (e.g.,  \citealp{deMedeiros2013,Reinhold2013,McQuillan2014,Canto2020}). One of the major challenges emerging from the data collected by these missions is the identification of rotation signatures in stars populating the intermediate and upper regions of the H-R diagram, namely, A-, B-, and O-type stars. From the analysis of the high-precision LCs produced by the {\em Kepler} mission, \citet{Balona2013} claimed the presence of photometric rotational periodicities in a sample of 875 A-type stars, on the basis of the Lomb-Scargle method (\citealp{Scargle1982,Horne1986,Press1989}), suggesting that 44\% of the total number of analyzed A-type stars present rotational modulation. In addition, \citet{Balona2016}, also on the basis of {\em Kepler} observations, suggested that rotational modulation in B stars is as common as it is among A stars. Nevertheless, a recent study \citep{Sikora2020} suggests that the incidence rate of inhomogeneous surface structures in A- and B-type main-sequence stars, potentially responsible for the observed rotational modulation, is less than approximately 30\%. A more recent study by \citet{Balona2019} analyzed the LCs of 160 B-type stars collected by the TESS space mission, most likely on the basis of the Lomb-Scargle technique, and concluded that, out of 114 main-sequence stars of the referred sample, 45 targets are rotational variables. A parallel study conducted by \citet{Pedersen2019}, dedicated to a search for variability in TESS O- and B-type stars, on the basis of asteroseismology, reported the detection of 21 rotational variables out of a sample of 154 stars. In addition, \citet{David2019} derived rotational periods for 13 previously known magnetic O, B, and A stars observed by TESS, performing a Lomb-Scargle analysis (\citealp{Lomb1976,Scargle1982}). In fact, these two latter studies explored targets in common with the stellar sample analyzed by \citet{Balona2019}. 

As underlined in the first paragraph of this section, the determination of rotation periodicities can reveal itself to be a complex task. In other words, to avoid misidentification of rotation signatures, such a task cannot be accomplished merely on the basis of only one analysis technique, as shown by different studies (\citealp{deMedeiros2013,Weingrill2015,Basri2018,Canto2020,Tan2020}). Of course, the Lomb-Scargle, Fast Fourrier Transform or autocorrelation function are appropriate methods to identify periodicities in stellar LCs but do not necessarily reflect traces of a phenomenon with astrophysical meaning. Readers are referred to \citet{deMedeiros2013}, \citet{Canto2020}, and \citet{Tan2020} for a consistent discussion about the complexity in the identification of stellar rotation signatures. 

This study contributes with an unprecedented manifold analysis of photometric LCs of B-type stars based on multiple techniques, namely, (i) Lomb-Scargle periodograms (e.g., \citealp{Scargle1982,Horne1986,Press1989}), (ii) the Fast Fourier Transform (FFT) (e.g., \citealp{Zhan2019}), and (iii) wavelet analysis (\citealp{Grossmann1984}). For the present purpose, we have chosen the sample of 160 B-type stars with LCs collected by the TESS space mission, analyzed by \citet{Balona2019}. The paper is organized as follows. Section~\ref{observation} presents the observational data set used and the analysis procedure applied in the search for variability signals and their classification. Section~\ref{sec:results} provides the main results along with an overall discussion and an examination of some particular cases, whereas a summary is presented in Section~\ref{sec:summary}.

\section{Stellar Sample and Data Analysis} \label{observation}

The TESS space mission \citep{Ricker2015} produces high-precision photometry of millions of stars. The primary mission plan is to survey almost the entire sky by monitoring 26 segments (or sectors) of 90$^o\times$24$^o$, each one with a duration of 27 days. The mission provides photometric data at different cadences, namely 2 and 30 minutes in Cycles 1 and 2 and 20 seconds, 2 minutes and 10 minutes in Cycle 3, with a time baseline from 27 days to 351 days, depending on sector overlaps. While 2-minute cadence data, also known as Target Pixel (TP) files, are available for a subset of targets, all CCDs, called full-frame images (FFIs), are read out every 30 minutes (10 minutes in the extended mission)\footnote{http://archive.stsci.edu/tess}. Subsets of TESS targets were observed for multiple sectors, with approximately 1-2\% of targets located in the Continuous Viewing Zone (CVZ) during the primary mission \citep{Barclay2018}, where targets were observed continuously for a year. Such a fact makes these targets particularly valuable for extracting long rotation periodicities and analyzing the persistence of stellar cycles. On July 4, 2020, TESS accomplished its primary mission, imaging approximately 75\% of the starry sky as part of a two-year-long survey. Now, in its extended mission, TESS is covering the southern hemisphere, which will be completed in September 2022. After spending a year resuming surveying of the southern sky, TESS will take another 15 months to collect additional observations in the north and to survey areas along the ecliptic not yet observed.

For the purpose of the present work, we chose 160 B-type stars with LCs collected by the TESS mission and previously analyzed by \citet{Balona2020} in the search for rotation periodicities probably using the Lomb-Scargle technique. Among these stars, 154 were analyzed by \citet{Pedersen2019} aiming to detect and classify variabilities using Discrete Fourier Transforms (DFT) \citep{Kurtz1985}. Nevertheless, our analysis provides two fundamental differences in relation to the referred studies. First, we apply a manifold analysis, using three different techniques: (i) Lomb-Scargle periodograms (e.g., \citealp{Scargle1982,Horne1986,Press1989}), (ii) Fast Fourier Transform (FFT) (e.g., \citealp{Zhan2019}), and (iii) wavelet analysis \citep{Grossmann1984}, accompanied by a visual inspection and a study of the position of the stars in the GAIA Color-Magnitude Diagram (GAIA CMD) for variable stars. Second, whereas the studies by \citet{Balona2020} and \citet{Pedersen2019} covered only observations from the first two sectors of the TESS mission, namely, Sectors~1 and~2, our work uses LCs of a significantly larger number of sectors, with 84 stars presenting LCs collected from at least 10 sectors, 32 of them observed in 15-16 sectors, 31 stars with more than 3 sectors, and 7 stars with observations in Sectors~1 and~2. Our analysis was applied for the 2-minute cadence LCs for all the sectors considered here, among sectors~1 to~29 observed from July~25, 2018 to September~22, 2020. These 2-min cadence data were downloaded from the {\em FFI-TP-LC-DV Bulk Downloads Page} of the Mikulski Archive for Space Telescopes\footnote{\url{https://archive.stsci.edu/tess/bulk_downloads.html}}. The TESS science processing operations center (SPOC) pipeline that produces the 2-min LCs is described by \citet{Jenkins2016}. The PDCSAP reduced LCs were used in the present study. An additional processing on these LCs was performed, when required, for the correction of possible distortions in the signature of periodicities, resulting from outlier removal and instrumental detrend, following the methods by \citet{deMedeiros2013}, \citet{Paz-Chinchon2015}, and \citet{Canto2020}, plus a removal of eventual transits following the strategy described in \citet{Paz-Chinchon2015}. Once such procedures were applied, we followed the same strategy defined by \citet{Bowman2018} to identify rotational signatures characteristic of hot stars and to determine their periodicities from the TESS LCs (see their Sect. 3.1). The entire sample of 160 stars is listed in Table~\ref{TABLE 1}, with relevant stellar parameters and the number of observed sectors taken into account in the present analysis. 

\begin{table*}[h]
   \centering
   \tablenum{1}
   \caption{\label{TABLE 1} List of 160 TESS B-type stars. The following information is listed: the TIC ID, effective temperature ($T_{eff}$) from \citet{Balona2019} (see their Table 2), GAIA magnitudes ($M_{G}$ and $G_{BP}-G_{RP}$) from \citet{Pedersen2019}, and TESS Sectors.}
    \begin{tabular}{ccrrl}
        \hline \hline
TIC       &  $T_{eff}$  & $M_{G}$ & $G_{BP}-G_{RP}$ &   TESS Sectors  \\
  & (K) & (mag) & (mag)   &    \\
\hline                
12359289	&	15330	&	-0.98	&	-0.12	&	 2\\
29990592	&	20665	&	-4.10	&	0.07	&	 1,2,3,4,5,6,7,8,10,11,12,13,27,28,29\\
30110048	&	17185	&	-2.02	&	-0.22	&	 1,2,3,4,5,6,7,8,10,11,12,13,27,28\\
30268695	&	22845	&	-2.20	&	-0.10	&	 1,2,3,4,5,6,7,9,10,11,12,13,27,28,29\\
30275662	&	17765	&	-3.37	&	-0.04	&	 1,2,3,4,5,6,7,8,10,11,12,13,27,28,29\\
...	&	...	&	...	&	...	&	 ...\\
\hline \hline                                              
    \end{tabular}
\tablecomments{Table 1 is published in its entirety in the machine-readable format. A portion is shown here for guidance regarding its form and content.} 
\end{table*}

\subsection{Why to apply a manifold procedure in the search for rotation periodicities from LCs? } \label{subsec:Analysis}

Once periodicities are identified, major challenges arise in classifying the signals as having physical meanings and then distinguishing rotation from pulsation, binarity or another variability phenomenon. This includes identifying how data series are uniformly sampled and how much the amplitude of the modulation is significant from the perspective of the associated error, the persistence of the rotation signature and the number of cycles observed in the time span, (e.g., \citealp{deMedeiros2013,Canto2020}), as well as the duration of intensity dips in the LC \citep{Tan2020}. Faced with this reality, the combination of the three techniques applied in our manifold analysis offers a solid path to overcome the referred challenges, complementing each other. For instance, the FFT method is efficient in the search for periodic signals that have high-duty cycles in uniformly sampled data series, whereas Lomb-Scargle is a powerful method, particularly for unevenly sampled data. In this sense, TESS LCs are nearly evenly sampled with a few irregularities, including some gaps. Sometimes those LCs require rebinning to regular time intervals close to their original bins and fulfilling eventual gaps with linear interpolation. The wavelet transform is a powerful technique to analyze a time series in the time-frequency domain, in other words, by decomposing periodicities as power spectra sections along the time window of the data. The wavelet maps can reveal in their detailed components particular characteristics that may not be evident in the time series themselves or in the global power spectra, including the persistence of the signals. As such, the wavelet method helps us in establishing the types of variability identified in an LC. We refer to \citet{Bravo2014} for a detailed analysis of different signatures that can be revealed in wavelet maps of stellar LCs. 

In this sense, the wavelet method is crucial to enrich the manifold analysis of an LC, revealing some peculiar patterns in the wavelet map that can help strongly in the identification of a variability type. For instance, the wavelet map can display how the amplitude of a signal may vary in time, with the presence of regular beats being a common characteristic of pulsators (e.g., \citealp{Canto2020}). Rotational modulation of cool stars typically presents semiregular flux variations caused by surface inhomogeneities and long-term amplitude variations associated with activity cycles (e.g., \citealp{Canto2020,Ferreira2015}).
In contrast, rotational modulation of chemically peculiar (CP) hot stars with surface inhomogeneities typically exhibits more regular flux variations, with possible amplitude variations that are considerably longer than in cool stars due to the presence of large-scale magnetic fields \citep{Bowman2018}. The amplitude of most CP stars are thus expected to be almost constant in a TESS LC time span (e.g., \citealp{Bowman2018}), a pattern that can be easily identified as a signature in the wavelet map. Some hot stars, CP or not, may, however, show amplitude variations with a similar signature to those of cool stars in the wavelet map (e.g., \citealp{Drury2017}).
Overall, the wavelet map also displays the persistence of a signal -- either with constant amplitude, regular beats or semiregular amplitude variations -- along the time span of an LC, thus helping in the establishment of a physical meaning to that signal. Finally, some stars may have a distribution of surface inhomogeneities such that their LCs display more than one dip per rotation cycle, usually two dips, whose signature is referred to as double-dip (\citealp{Basri2018,Basri2018b}). Such a signature can be verified from visual inspection of the LC, and it can be clearly identified in a wavelet map from the presence of two persisting features with integer multiple periods along the time span. This is also an important aspect to be considered for a more reliable determination of the actual rotation period of a star (e.g., \citealp{Bowman2018,Canto2020}). While the identification of rotational signatures of hot stars was based on the criteria described by \citet{Bowman2018}, in practice, all the criteria described above were considered by following the approach used by \citet{Pedersen2019}. As such, a visual inspection of the LCs and an analysis of periodicity spectra was conducted by several of the authors independently. For those stars presenting additional eclipsing binary (EB) or pulsation features, we prewhitened the referred signatures from the LCs and searched for rotation periodicity in the residual LCs. This was particularly applied for the stars TIC 150442264, TIC 350823719, and TIC 469906369.

\section{Results} \label{sec:results}

We performed a manifold analysis on the LCs of 160 TESS B-type targets, first dedicated to the search for rotational variability. Periodicities were computed for the targets with unambiguous rotational signatures, along with an investigation of the signature persistence. We also computed the phase-folded LCs. In fact, differences in the phased LCs associated with fluctuations in the O-C diagrams, a power excess in the residual frequency spectra, are of great importance in the classification of the class of variability beyond rotation, pointing not only to intrinsic instabilities in the nature of the variability but also to the presence of instrumental effects or contamination from sources that may vary between sectors. The main results are presented in detail in the following subsections.

\begin{table*}[h]
    \centering
   \tablenum{2}
   \caption{\label{TABLE 2} TESS B-type stars with rotation periodicity from the combined analysis of FFT, Lomb-Scargle, and wavelet. The following information is listed: TIC ID, rotation period ($P_{rot}^{'}$) from \citet{Balona2019}, rotation period revealed by the present work ($P_{rot}$), where the number between parentheses corresponds to the uncertainty on the final digit of the reported period, level of signature persistence (LSP), effective time span ($t_{SPAN}$) and effective number of cycles ($N_{Cycle}$) obtained from our analysis, classifications from \citet{Balona2019} and \citet{Pedersen2019}, and Remarks from our analysis ($a$ - Rotation+EB, $b$ - Rotation+Multi Flares, and $c$ - Rotation+Pulsation). The parameter $P_{rot}^{'}$ was obtained from the frequency $\nu_{rot}$ given by \citet{Balona2019}.}
    \begin{tabular}{rcccccllc}
        \hline \hline
TIC	&	$P_{rot}^{'}$	&	$P_{rot}$	& 	LSP	&	$t_{SPAN}$	&	$N_{Cycle}$	& \cite{Balona2019} &	\citet{Pedersen2019}	&	Remarks\\
	&	(days)	&	(days)	&	(\%)	&	(days)	&		&&		&	\\
\hline
12359289	&	3.077	&	3.065(4)&		96	&	25	&	8.4	&SXARI&	rot	&	\\
38602305	&	2.976 &	2.9686(4)&		97	&	362	&	126.0	&ROT&	rot 	&	 \\
41331819	&	1.401	&	1.40032(9)&		93	&	266	&	195.4	&ROT&	rot/outburst?	&	\\
89545031	&	3.759	&	3.75413(2)&		100	&	46	&	12.7	&SXARI (ACV)&	rot	&	\\
139468902	&	0.455	&	0.45707(7)&		99	&	45	&	102.0	&ROT&	rot/SPB?	&	\\
141281495	&	2.994	&	3.0086(2)&		86	&	319	&	110.1	&ROT&	rot?	&	\\
149039372	&	--	&	0.322260(5)&		100	&	317	&	1042.8	&SPB?&	rot?/SPB?	&	\\
150442264	&	--	&	1.490911(7)&		98	&	335	&	232.0	&EB (EB:)&	EB+puls/rot	&	$a$\\
182909257	&	3.135	&	3.1390(3)&		97	&	52	&	17.1	&SXARI&	rot	&	\\
197641601	&	3.436	&	4.987(2)&		79	&	44	&	8.9	&ROT?&	instr/rot 	&	  \\
224244458	&	1.916	&	1.9311(5)&		93	&	46	&	24.6	&SXARI+FLARE&	rot+mini-outburst?	&	$b$\\
231122278	&	--	&	4.549(4)&		87	&	340	&	76.7	&SPB&	rot/SPB?	&	\\
262815962	&	2.710	&	2.733(8)&		84	&	44	&	16.6	&ROT&	rot/SPB?	&	\\
270070443	&	2.532	&	2.535(6)&		99	&	25	&	10.2	&SXARI&	rot	&	\\
279430029	&	0.561	&	0.56085(3)&		49	&	357	&	655.0	&ROT&	rot/Be	&	\\
279511712	&	1.653	&	1.65167(0)&		100	&	338	&	211.3&ELL (LPB)&	rot	&	\\
280051467	&	4.367	&	4.3720(5)&		98	&	122	&	28.8	&SXARI&	rot	&	\\
294747615	&	5.208	&	5.1918(5)&		97	&	344	&	68.4	&SXARI&	rot/SLF?	&	\\
300744369	&	1.045	&	1.04428(9)&	100	&	340	&	336.3	&ROT&	rot	&	\\
307291308	&	2.584	&	1.29641(4)&		97	&	233	&	185.1	&SXARI&	instr/rot?	&	\\
313934087	&	--	&	1.75222(5)&		84	&	45	&	27.2	&SPB&	SPB/rot	&	\\
350146577	&	1.838	&	1.837527(3)&		100	&	300	&	168.4	&SXARI&	rot	&	\\
350823719	&	--	&	4.1963(9)&		97	&	360	&	88.9	&EB&	SPB?/rot	& $c$\\
354671857	&	0.344	&	0.343705(9)&		100	&	66	&	198.2	&ROT?&	SPB?/rot	&	\\
355653322	&	0.790	&	0.79318(8)&		88	&	46	&	60.0	&ROT&	rot?/outburst?/instr?	&	\\
410447919	&	4.566	&	4.2331(4)&		85	&	247	&	60.8	&ROT?&	rot/SPB?	&	\\
410451677	&	2.049	&	2.05517(6)&	98	&	176	&	88.4	&SXARI&	rot	&	\\
469906369	&	--	&	2.8535(1)&		69	&	68	&	24.8	&MAIA&	SPB/$\beta$Cep/instr?	&	$c$\\\hline 
\hline
    \end{tabular}
\end{table*}

\subsection{Rotational signatures} \label{subsec:rotation}

A summary of the B-type TESS stars with rotational signatures issued from our analysis is presented in Table~\ref{TABLE 2}, corresponding to 28 stars out of the entire sample of 160 targets, for which we computed confident periodicities on the basis of our manifold procedure. The rotational periods listed in the referred table are those computed from the Lomb-Scargle periodograms, which are consistent with those obtained from FFT and wavelet analysis. The uncertainties on the periods were estimated according to the formal error definition given by David-Uraz et al. (2019). Figure Set 1 in the Appendix displays for each rotational variable star the TESS LC, with the corresponding phase-folded LC, the FFT and Lomb-Scargle frequency spectra, and the wavelet maps. Among these 28 rotational variables, 26 present a level of persistence of rotational signature greater than or equal to 79\%, and only 2 stars show a persistence between 49 and 69\%. In fact, following \citet{deMedeiros2013} and \citet{Canto2020}, we consider that stars have confident rotation periods when their LCs exhibit more than three observed cycles, where the effective number of cycles ($N_{Cycle}$) is the effective time span of the observation ($t_{SPAN}$) of the LC, excluding gaps, divided by the rotation period ($P_{rot}$). Rotation periods were computed for six TESS B-type stars, namely, TIC 149039372, TIC 150442264, TIC 231122278, TIC 313934087, TIC 350823719, and TIC 469906369, with periodicities not yet reported in the literature. The following stars have rotation periods previously given by \citet{David2019}: TIC 89545031  ($P_{rot} = 3.7349 \pm 0.0005$) and TIC 279511712 ($P_{rot} = 1.65183 \pm 0.00002$), and also given by \citet{Cunha2019}: TIC 12359289 ($P_{rot} = 3.06395 \pm 0.00041$), TIC 89545031 ($P_{rot} = 3.72251 \pm 0.00097$), TIC 182909257 ($P_{rot} = 3.14108 \pm 0.00082$), and TIC 350146577 ($P_{rot} = 1.83764 \pm 0.00004$) which, within the uncertainties, are consistent with our determinations. A few stars listed in Table~\ref{TABLE 2} deserve special attention due to their short rotation periodicities, specifically those with periods shorter than approximately 0.5 days. These rotational variables, in particular TIC 139468902, TIC 149039372, and TIC 354671857 may belong to the known class of fast-rotating B-type stars (\citealp{Mowlavi2016,Balona2016}).
To a consistency check of the rotation periodicities listed in Table~\ref{TABLE 2}, we followed the methodology described in \citet{Kochukhov2021}. According to those authors, periods with corresponding ratios of the equatorial to the breakup velocity close to or less than 1 may be consistent with rotation. The stars TIC 139468902, TIC 149039372, TIC 279430029, TIC 354671857, and TIC 355633322, all with $P_{rot} < 1$, are fully compatible with those criteria and have clear signatures of rotational modulation. For TIC 354671857, \citet{Kochukhov2021} give a rotation period of 0.344~days, which agrees with the value obtained in the present work.

\begin{table*}[h]
    \centering
   \tablenum{3}
   \caption{\label{TABLE 3} TESS B-type stars with noisy behavior from the present manifold analysis. The following information is listed: TIC ID, possible frequency $\nu_{rot}$ given by \citet{Balona2019}, classifications from \citet{Balona2019} and \citet{Pedersen2019}, and Remarks from our analysis.}
    \begin{tabular}{rclll}
        \hline \hline
TIC	&	$\nu_{rot}$	&\cite{Balona2019} &	\citet{Pedersen2019}&	Remarks	\\
	&	(d$^{-1}$)	&	&	&		\\
\hline							
31674330	&	--	&  -- &	--	&	Noisy	\\
33945685	&	2.835	& ROT&	instr?($\nu$inst'2.8d$^{-1}$)/plus	&	Noisy	\\
49687057	&	--	&  -- &	instr/binary?	&	Noisy	\\
147283842	&	--	&  -- &	--	&	Noisy+Ambiguous variability	\\
152283270	&	0.434	& EA+ROT?&	instr/binary(transit)	&	Noisy	\\
167415960	&	--	&  -- &	const?	&		Noisy\\
176935619	&	2.747	& ROT&	instr?($\nu$inst'2.8d$^{-1}$)	&		Noisy\\
229013861	&	0.423	& ROT?&	rot	&	Noisy	\\
260368525	&	1.193	& ROT?&	SPB?	&	Noisy	\\
260540898	&	0.562	& ROT?&	rot?	&	Noisy	\\
260820871	&	--	& EP&	rot/binary?	&	Noisy	\\
270557257	&	2.475	& ROT?&	instr($\nu$inst'2.8d$^{-1}$)	&	Noisy	\\
278683664	&	1.701	& ROT?&	const??	&	Noisy	\\
278865766	&	--	&  -- &	const??	&	Noisy	\\
278867172	&	--	&  -- &	const??	&		Noisy\\
300325379	&	1.838	& ROT&	rot?	&		Noisy\\
306672432	&	1.091&	ROT?&	const?/rot?	&	Noisy	\\
308454245	&	--	& MAIA&	$\delta$Sct	&	Noisy	\\
308456810	&	0.251	& ROT&	rot?	&	Noisy/Ambiguous variability	\\
308748912	&	--	&  -- &	SLF?/outburst?/instr	&		Noisy\\
349829477	&	--	&  -- &	const?	&	Noisy	\\
355141264	&	--	&  -- 	&const?	&	Noisy	\\
355477670	&	0.272	& ROT?&	const?	&		Noisy\\
364421326	&	1.976	&ROT?&	rot?	&		Noisy\\
369397090	&	--	&  -- &	--	&	Noisy	\\
370038084	&	--	& -- &	const?	&		Noisy\\
372913233	&	--	&  -- &	const?/outburst?/instr?	&	Noisy	\\
372913582	&	--	& 	SPB?&const?/outburst?/instr?	&	Noisy	\\
441196602	&	0.144	&ROT?&	const?	&	Noisy	\\
\hline \hline                                            	  
    \end{tabular}
\end{table*}

However, some major aspects emerge from the present analysis when the bulk of our results is compared with the recent literature. First, the 29 stars listed in Table~\ref{TABLE 3} exhibit a clear noisy behavior in their LCs, from which our manifold analysis shows no periodicities with physical meaning. Such a finding is completely at odds with \citet{Balona2019}, who classified 14 of these stars as rotational variables, computing their rotational frequencies. Table~\ref{TABLE 3} also presents discrepancies with \citet{Pedersen2019}, with a few stars classified by these authors as possible rotation variables showing rather noisy behavior. Based on \citet{deMedeiros2013}, we consider a noisy LC when the ratio of the signal amplitude to a noise level, estimated from the standard deviation of the residual, is less than a threshold. According to a careful analysis of the current sample, we identified a value of 0.2 as producing a reasonable separation between noisy LCs and those with physical signals. Second, among the stars listed in Table~\ref{TABLE 4}, 46 show rather a pulsational, binarity or ambiguous variability behavior in their LCs, with no clear rotation signatures, also in contrast with previous classifications (\citealp{Balona2019,Pedersen2019}), that suggested for them a possible rotational variability. Following \citet{Canto2020}, we define as ambiguous variability those stars showing visually noticeable fluctuations in the LCs, but faint enough for a reliable interpretation or with insufficient time span for proper identification of the variability nature, as well as significant large-amplitude variations but presenting an irregular or complex behavior sometimes also caused by systematics. 

\begin{table*}[h]
    \centering
   \tablenum{4}
   \caption{\label{TABLE 4} TESS B-type stars with ambiguous variability from the present manifold analysis. The following information is listed: TIC ID, possible frequency $\nu_{rot}$ given by \citet{Balona2019}, classifications from \citet{Balona2019} and \citet{Pedersen2019}, and Remarks from our analysis.}
    \begin{tabular}{rclll}
        \hline \hline
TIC      	&	 $\nu_{rot}$   & \citet{Balona2019}	&	 \citet{Pedersen2019}	&	Remarks	\\
 	&	  (d$^{-1}$) 	&	 &	&	               	\\
\hline							
30317301	&	--	&EB&	EV/rot/SLF	&	Ambiguous variability/EB/Pulsation	\\
31867144	&	--	&SPB&	rot/SPB+$\beta$Cephybrid?	&	Ambiguous variability/Rotation+SPB	\\
47296054	&	0.836	& SPB+ROT&	rot/SPB/Be	&	Ambiguous variability/Rotation+Pulsation	\\
53992511	&	--	& BE(BE)&	rot/SPB/Be	&	Ambiguous variability/Rotation/SPB	\\
55295028	&	--	&BE&	rot/SPB/Be	&	Ambiguous variability/Pulsation+Rot	\\
66497441	&	--		& ELL&	EV/rot?/SPB?	&	SPB	\\
92136299	&	2.252	& ROT+FLARE?&	SPB+Be-type mini-outbursts/rot	&	SPB+Flares\\
115177591	&	1.736	& 	ROT&SPB 	&	SPB   	\\
118327563	&	4.367	& ROT+FLARE?&	--  	&	Ambiguous variability/sdB Pulsation?   	\\
150357404	&	0.640	& SPB+ROT?&	SPB 	&	SPB   	\\
167045028	&	0.329	& EB?&	rot	&	Ambiguous variability/Rotation?/EB?	\\
169285097	&	--	& sdB hybrid&	--	&		Ambiguous variability/Pulsation\\
176955379	&	2.994	& ROT&	SPB/rot   	&	Ambiguous variability/Pulsation+Rot   	\\
177075997	&	0.352	& ROT?&	instr / rot?	&	Ambiguous variability	\\
206547467	&	0.233	& ROT&	rot/const?/SPB? 	&	Ambiguous variability   	\\
207235278	&	0.558	& ELL&	EV/rot    	&	Binary  	\\
230981971	&	--	& BE(BE)&	rot/SPB?/Be	&	Ambiguous variability	\\
238194921	&	0.727	& SPB+ROT&	rot/SPB   	&	Ambiguous variability 	\\
259862349	&	0.402	& ROT?&	instr	&	Ambiguous variability/Rotation	\\
261205462	&	--		& SPB&	SPB/rot	&	Ambiguous variability/SPB?/Rotation?	\\
271503441	&	3.058	& ROT&	SPB?/outburst?	&	Ambiguous variability/SPB?\\
271971626	&	0.215	& ROT+MAIA&	rot	&	Ambiguous variability/Rotation?/EB?	\\
277022967	&	--	& ACYG&	rot/SPB?/SLF?/Be	&	Ambiguous variability/Rotation/IGW?	\\
277103567	&	1.497	& ROT(BE)&	rot/EV?	&	Ambiguous variability/Pulsation+Rot	\\
277982164	&	1.266	& ROT&	rot/SPB	&	Ambiguous variability/Rotation+Pulsation	\\
279957111	&	--	&  -- &	rot	&	Ambiguous variability/Pulsation	\\
280684074	&	0.563	& SPB+ROT(LPB)&	SPB 	&	SPB 	\\
281741629	&	--	& BE(BE)&	rot/Be	&	SPB?	\\
293268667	&	3.390	& SPB+ROT&	SPB/rot?   	&	SPB	\\
293973218	&	--	& SPB&	rot/SPB	&	Ambiguous variability	\\
300010961	&	0.683	& ROT+MAIA&	rot/$\beta$Cep?	&	$\beta$ Cep+Oscillation?	\\
300329728	&	0.411	& ELL/ROT&	rot	&	Ambiguous variability/$\beta$ Cep?/EB?	\\
300865934	&	--	&  -- &	instr/outburst?	&		Ambiguous variability\\
306824672	&	--	& SPB?&	rot?/SPB?	&	Ambiguous variability	\\
307291318	&	--		&  -- &	instr/rot?	&	Ambiguous variability/Rotation+Transity?	\\
307993483	&	3.709	& SPB+ROT?&	$\beta$Cep?/SPB 	&	Ambiguous variability+Flares   	\\
308537791	&	0.527	& ROT+MAIA&	rot	&	Pulsation/Oscillation	\\
327856894	&	--	&  -- &	rot?/outburst?/instr?	&	Ambiguous variability	\\
358466708	&	1.661	& ROT&	rot/SPB   	&	SPB 	\\
358467049	&	0.265	& SXARI&	rot	&	SPB	\\
358467087	&	--	& SPB?&	SPB/EV?/rot?/SLF?/instr?	&	Ambiguous variability	\\
364323837	&	--& SPB?	&	rot?/SPB?	&	Ambiguous variability/Pulsation	\\
364398190	&	0.760	& ROT&	rot?/SLF? 	&	Ambiguous variability   	\\
364398342	&	--	& BE(GCAS)&	rot?/SPB?/Be	&	Ambiguous variability	\\
369457005	&	--	&	SPB?&	rot/SPB?	&	Ambiguous variability/SPB or SPB+Rot\\
372913684	&	1.460	&	SXARI (EA:)&	rot	&	Ambiguous variability/Rotation+EB\\
373843852	&	--	& ACYG&	SPB?/rot?	&	Ambiguous variability/IGW	\\
425057879	&	--	&	EB/ELL?&	instr?/binary?/rot?	&	Ambiguous variability\\\hline \hline
    \end{tabular}
\end{table*}

\begin{figure}[h!]
	\centering
	\includegraphics[scale=.6]{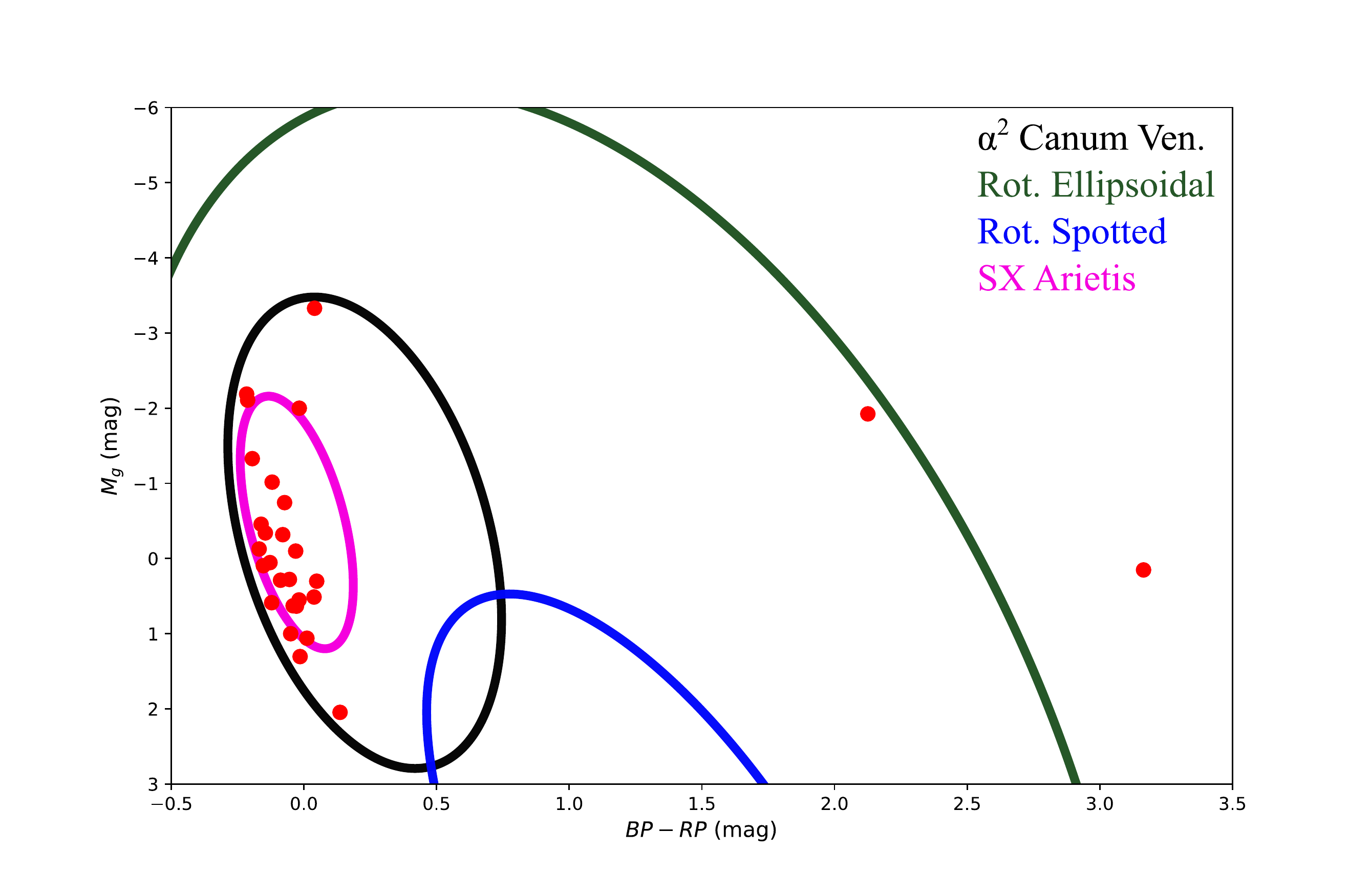}
	\caption{GAIA Color--Magnitude Diagram (CMD) for the variable stars analyzed in this study, presenting rotation signatures. Ellipses were constructed from the scenario presented in Fig. 4 of \citet{Gaia2019} for GAIA rotation-induced variables.} 
	\label{HR}
\end{figure}

\subsection{Pulsation signatures and ambiguous variability} \label{subsec:pulsation}

The sample of 160 targets analyzed by \citet{Balona2019} and \citet{Pedersen2019} for variability classification was based on observations collected by TESS in Sectors~1 (July 25 -- August 22, 2018) and~2 (August 23 -- September 20, 2018). Because the present-day TESS database offers new observations for 123 stars in a large number of sectors, out of the sample of 160 targets, we revisited the referred classifications. The additional TESS Sectors are listed in Table~\ref{TABLE 1}, from which one can see that 84 stars were observed in at least 10 TESS Sectors. Indeed, the detection of stellar photometric variability depends strongly on not only instrumental characteristics, such as photometric sensitivity, but even on the time span of the observation (e.g., \citealp{Leao2015,Canto2020}) and on LC reduction and post-treatment procedures (\citealp{Bravo2014,Lira2019,Canto2020}). Special emphasis was placed on the stars classified in previous studies as rotational variables. Tables~\ref{TABLE 2}, \ref{TABLE 3}, and~\ref{TABLE 4} list the classification issued from the present analysis and those from \citet{Pedersen2019} and \citet{Balona2019}. In the following, we present the major constraints that are perhaps contributing to important discrepancies among the referred classifications. 

\subsection{Astrophysical implications} \label{subsec:astrophys}

The results emerging from our analysis have relevant astrophysical implications associated with the treatment procedures adopted in the computation of periodicities and with the variability classification for the sample analyzed here. The first implication arises from the disagreement between the variability classification issued from the present work and those by \citet{Balona2019} and \citet{Pedersen2019}, which can be understood in two ways. First, those authors based their analyses on the LCs acquired in the first two sectors of the TESS mission, namely, Sectors~1 and~2. In contrast, our analysis covers a significantly larger number of sectors, with 84 stars observed in at least 10 sectors, 32 of which are observed along 15-16 sectors. Considering that the time range of observations is a crucial aspect for a solid variability classification, the use of only two sectors can statistically impoverish the analysis. Second, the referred studies applied a single analysis technique, whereas our diagnostic in the identification of variability signatures was based on a manifold procedure derived from Lomb-Scargle, FFT, and wavelet analysis, which is even able to identify the persistence associated with variability phenomena. These two aspects also lie at the core of the misidentification of rotational periodicities and variable classes, such as those extracted from noisy LCs. In this context, a comparison of our periodicities with those from  \citet{Balona2019} shows the following portrait: (i) for 19 stars listed in Table~\ref{TABLE 2}, we find an excellent agreement with \citet{Balona2019}, with a root mean-square of the differences of 0.009 (these 19 stars have essentially very well defined LCs); (ii) for the stars TIC 197641601, TIC 307291308, and TIC 410447919, also listed in Table~\ref{TABLE 2}, we did not find periodicities compatible with those given by \citet{Balona2019}; and (iii) for the 14 stars listed in Table~\ref{TABLE 3} with frequencies computed by \citet{Balona2019}, for which we have identified rather noisy behavior in their LCs, we did not find periodicities with a clear physical meaning. Regarding point~(ii), the difference in $P_{rot}$ for TIC 197641601 came likely from an instrumental anomaly observed in sector 1, also pointed by \citet{Pedersen2019}. For TIC 307291308, and TIC 410447919, the differences came, in fact, from the larger number of sectors considered in the present analysis.

The second astrophysical implication derived from the current results come out from Fig.~\ref{HR}, which shows an overview of the classes of rotational variables identified in the present study. This figure displays the stars listed in Table~\ref{TABLE 2} in a GAIA second data release~(DR2) Color-Magnitude Diagram~(CMD), with ellipses enclosuring different classes of rotational variables, based on Fig.~4 of \citet{Gaia2019}. The scenario emerging from the referred CMD points for different classes of rotational variables, composed at least of $\alpha^2$ Canum Venaticorum, rotating ellipsoidal, and SX Arietis stars. The first class consists of highly magnetic variable Bp- and Ap-type main sequence stars. The SX Arietis variables are similar to $\alpha^2$ Canum Venaticorum stars but with higher temperatures, and, by consequence, their location in the CMD presents some overlap of the two distributions for these two variable types. The rotating ellipsoidal variables are close binary systems with ellipsoidal components, which show variability (without eclipses) due to orbital motion of a star that is distorted by the other component of the system. Overall, rotational modulation in B-type stars is usually interpreted as being due to temperature and/or chemical spots on the stellar surface caused by large-scale magnetic fields. Nevertheless, the formation mechanism of these magnetic fields and their evolution are still a matter of much debate (e.g., \citealp{Schneider2016,Villebrun2019,Landstreet2007,Landstreet2008,Fossati2016,Shultz2019}). The main challenge to answering these questions seems to be hampered by the paucity of present-day statistics. For instance, the literature reports fewer than 100 known magnetic early B-type stars \citep{Shultz2018}. In this context, the B-type rotational variable candidates revealed by TESS, as the $\alpha^2$ Canum Venaticorum and SX Arietis types constrained in Fig. \ref{HR}, may represent additional laboratories for direct or indirect magnetic diagnostics. Interestingly, the stars listed in Table~\ref{TABLE 2} show $P_{rot}$ values between 0.322260 and 5.1918 days, a range of values compatible with the large majority of rotation periodicities of magnetic B-type stars reported by \citet{Shultz2018} and \citet{Bowman2018}. These studies show that the referred stars have rotation periodicities predominantly lower than approximately 6 d.

\section{Summary} \label{sec:summary}

In this work, we report the results of a manifold analysis of TESS LCs with a 2-min cadence for 160 B-type stars, using a considerably large number of sectors in comparison with previous studies. The analyzed sample includes 83 stars previously classified as possible rotational variables. The main findings from this analysis can be summarized as follows.

i) We computed rotational periodicities for 6 new TESS B-type stars.

ii) For 14 stars with previously computed rotational periods, we identified rather noisy behavior in their LCs.

iii) For 46 stars, out of the sample of 83 TESS B-type stars previously classified as possible rotational variables, our analysis reveals pulsational, binarity or ambiguous variability behavior rather than rotation signatures.

iv) The total sample of 28 potential rotators analyzed in the present study shows an overlap of different classes of rotational variables composed of $\alpha^2$ Canum Venaticorum, rotating ellipsoidal, and SX Arietis stars.

Magnetic hot stars are endowed with several physical processes in their interiors, including rotation, pulsation, and the production of large-scale magnetic fields. The interaction between those processes yields in large uncertainties to theoretical models of stellar structure and evolution for the upper main-sequence region (e.g, \citealp{Bowman2018}). For instance, theoretical models predict that large-scale magnetic fields induce uniform rotation and may produce smaller convective-core overshooting region in those stars (e.g., \citealp{Browning2004}), even though more observational constraints is necessary to refine those predictions. Besides, asteroseismic theoretical models of hot stars predict that those magnetic fields and rotation may impact substantially in the excitation of waves (e.g., \citealp{Lecoanet2017,Mathis2014}). Therefore, the sample of the present work brings relevant constraints to improve the evolutionary models, especially for those considering rotation (e.g., \citealp{Maeder2009}), being also of significant interest for improving, theoretically and observationally, asteroseismic studies of those stars.

Some clear variabilities, classified here as ambiguous variabilities, that point to a rather difficult discrimination among rotation, pulsation or other signatures, should be revisited in future works, especially using additional observations with significantly larger time spans. The present findings reinforce previous studies carried out by different authors, showing that the detection of stellar rotation variability depends strongly on instrumental characteristics, such as photometric sensitivity, time span of the observation, LC reduction and post-treatment procedures, and even on the number, size, and location of spots along the stellar surface. In fact, the inference of stellar surface properties from photometric LCs is clearly an \textit{ill-posed problem} because the solutions clearly depend discontinuously upon the initial modeling assumptions (e.g., \citealp{Luger2021}). In the same direction, the present study reinforces that the extraction of rotation periodicities from observed stellar LCs appears to be a \textit{non-well posed problem} \citep{Hadamard1902} in the sense that the solution may depend upon different conditions, as underlined above. 

\acknowledgments

We warmly thank our families for attending us with care, patience, and tenderness during the home office tasks for the preparation of this study in the face of this difficult COVID-19 moment. Research activities of the observational astronomy board at the Federal University of Rio Grande do Norte are supported by continuous grants from the Brazilian funding agencies CNPq, FAPERN, and INCT-INEspa\c{c}o. This study was financed in part by the Coordena\c{c}\~ao de Aperfei\c{c}oamento de Pessoal de N\'ivel Superior - Brasil (CAPES) - Finance Code 001. LFB acknowledges the UFRN/CNPq undergraduate fellowship. RLG and YSM acknowledge CAPES graduate fellowships, and MAT acknowledges the CNPq graduate fellowship. LAA acknowledges a CAPES/PNPD fellowship. BLCM, EJP, ICL, and JRM acknowledge CNPq research fellowships. This work includes data collected by the TESS mission. We acknowledge the use of public TIC Release data from pipelines at the TESS Science Office and at the TESS Science Processing Operations Center. TESS data were obtained from the Mikulski Archive for Space Telescopes (MAST). Funding for the TESS mission is provided by NASA's Science Mission directorate. We warmly thank the Referee for comments and suggestions that clarified important aspects of this study.

\appendix
\section{Online Material}

\begin{figure}[h!]
	\centering
	\includegraphics[scale=.6]{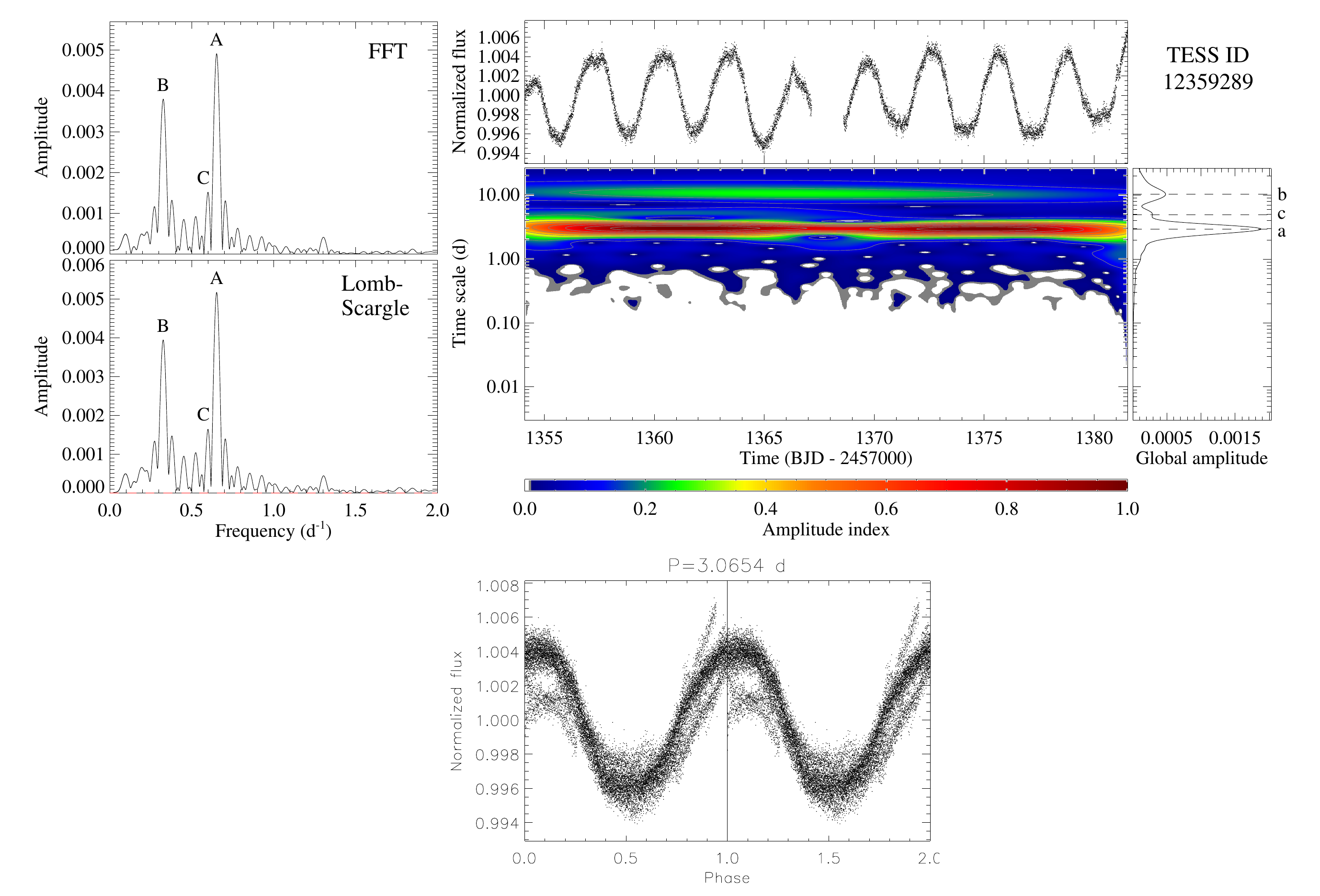}
	\caption{TESS LCs, LC phase folded, FFT and Lomb-Scargle periodograms, and wavelet maps for TIC~12359289. The complete figure set (28 images) is available in the online Journal.} 
	\label{suppfig}
\end{figure}

Supplementary material supporting this study is also available
    (Figure~\ref{suppfig}), including post-treated light curves, wavelet
    maps, and periodograms, for stars with unambiguous rotation signatures. 

\figsetstart
\figsetnum{1}
\figsettitle{Diagnostic plots for the 28 TOIs with unambiguous rotation periods.}

\figsetend



\end{document}